\documentclass[superscriptaddress,aps,prl,twocolumn,showpacs,preprintnumbers,amsmath,amssymb,floatfix,10pt]{revtex4-1}
\usepackage{graphicx}
\usepackage{dcolumn}
\usepackage{bm}

\def\e{\epsilon}

\def\e{\mathcal{E}}

\def\ba{\begin{eqnarray}}
\def\ea{\end{eqnarray}}
\def\beq{\begin{equation}}
\def\eeq{\end{equation}}

\begin{document}

\title{Photon-Photon Interactions via Rydberg Blockade}

\author{Alexey V. Gorshkov}
\affiliation{Institute for Quantum Information, California Institute of Technology, Pasadena, California 91125, USA}
\author{Johannes Otterbach}
\affiliation{Department of Physics and Research Center OPTIMAS, Technische Universit\"at Kaiserslautern, 67663, Kaiserslautern, Germany}
\author{Michael Fleischhauer}
\affiliation{Department of Physics and Research Center OPTIMAS, Technische Universit\"at Kaiserslautern, 67663, Kaiserslautern, Germany}
\author{Thomas Pohl}
\affiliation{Max Planck Institute for the Physics of Complex Systems, 01187 Dresden, Germany}
\author{Mikhail D. Lukin}
\affiliation{Physics Department, Harvard University, Cambridge, Massachusetts 02138, USA}

\date{\today}

\begin{abstract}

We develop the theory of light propagation under the conditions of electromagnetically induced transparency (EIT) in systems involving strongly interacting Rydberg states. Taking into account the quantum nature and the spatial propagation of light, we analyze interactions involving few-photon pulses. We demonstrate that this system can be used for the generation of nonclassical states of light including trains of single photons with an avoided volume between them, for implementing photon-photon quantum gates, as well as for studying many-body phenomena with strongly correlated photons.

\end{abstract}

%42.50.Nn	Quantum optical phenomena in absorbing, amplifying, dispersive and conducting media; cooperative phenomena in quantum optical systems
%42.50.Gy	Effects of atomic coherence on propagation, absorption, and amplification of light; electromagnetically induced transparency and absorption
%32.80.Ee	Rydberg states
%32.80.Rm	Multiphoton ionization and excitation to highly excited states
%34.20.Cf	Interatomic potentials and forces

\maketitle

%for cutting, can put together michael's 3 perturbative regimes (and group other)
%for cutting, can remove the sentence "R enters only through boundary conditions..."
%for cutting, play with spaces in the definitions of counterpropagting amplitudes
%for putting back: if r-r interaction is negligible, applies to coherent pulses
%for putting back: my counterpropagating resonant paragraph about timing information, the original dark state polaritons, and a heralded single-photon source (if r-r interactions strong, then first get copropagating and reduce to single photon, and then these two single photons scatter) - we do make a claim of this sort in the intro, so maybe should put back into text later

%\textit{Introduction and basic intuition.}---
The phenomenon of electromagnetically induced transparency (EIT) \cite{fleischhauer05} in systems involving Rydberg states
\cite{saffman10} has recently attracted significant experimental
\cite{mohapatra07,mauger07,mohapatra08,zhao09b,kubler10,tauschinsky10,schempp10,pritchard10}
and theoretical \cite{lukin01b,friedler05,petrosyan08,moller08,muller09,weimer10,pohl10,olmos10,shahmoon10,ates11,honer11} attention.
While EIT allows for strong atom-light interactions without absorption, Rydberg states provide strong
long-range atom-atom interactions. Therefore, the resulting combination of EIT with Rydberg atoms is ideal
for implementing mesoscopic quantum gates \cite{saffman10,muller09} and 
for inducing strong photon-photon interactions, with applications
%with applications to the creation of non-classical states of light \cite{pohl10} including single-photon states \cite{lukin01b,saffman02porras08pedersen09,olmos10,honer11}, 
to photonic quantum information processing \cite{lukin01b,saffman02porras08pedersen09,pohl10,olmos10,honer11,friedler05,petrosyan08,shahmoon10,saffman10} and to the realization of
many-body phenomena with strongly interacting photons \cite{chang08}.
At the same time, the many-body theoretical description 
%Due to its %intrinsically 
%complicated many-body nature, the theory 
of EIT with arbitrarily strongly interacting Rydberg atoms, %with arbitrarily strong interactions, 
taking into  account the full quantum dynamics and the spatial propagation of light, has not been reported previously. 

In this Letter, we develop such a theory by analyzing  
%we make the first steps towards developing such a theory by solving 
the problem for at most two incident 
photons, which, in turn, provides intuition for understanding the full multi-photon problem.
We show that Rydberg atom interactions give rise to photon-photon interactions, which, below a critical inter-photonic distance, 
turn the EIT medium into an effective two-level medium. This can be used to implement photon-atom 
%as well as 
and photon-photon phase gates and to enable deterministic single-photon sources.

%%%%%%%%%%%%%%%%%%%%%%%%%%%%%%%%%%%%%%%%%%%%%%%%%%%%

\begin{figure}[b!]
\begin{center}
\includegraphics[width = \columnwidth]{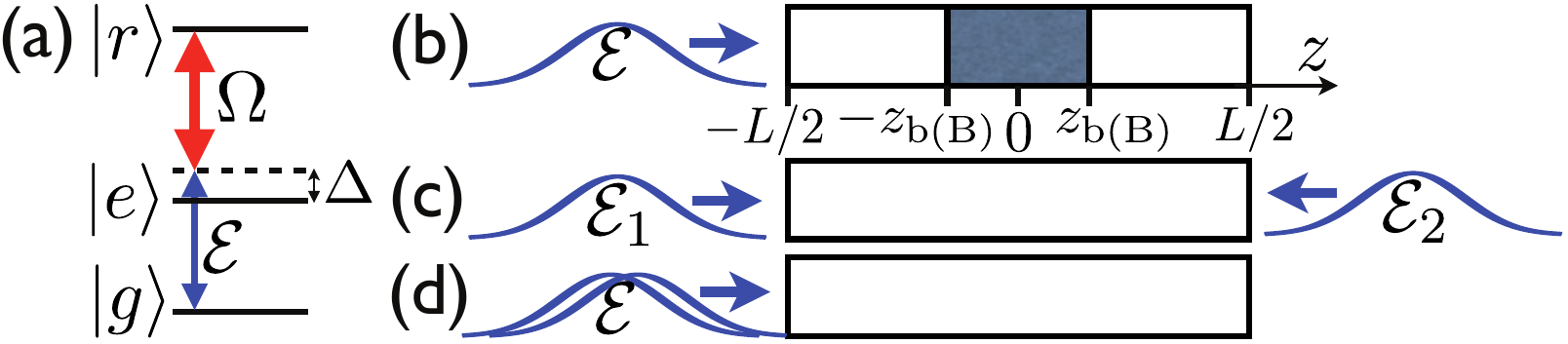}
\caption{(a) EIT level scheme, in which a %atomic 
ground state $|g\rangle$, %low-lying 
an excited state $|e\rangle$, and a strongly interacting Rydberg state $|r\rangle$ are coupled by a %propagating 
quantum probe field $\e$ and a %strong 
classical control field with Rabi frequency $\Omega$ and single-photon detuning $\Delta$. (b) Interaction of one photon with a Rydberg excitation stored at $z = 0$, which modifies the propagation within the blockade region $|z|<z_{\rm b(B)}$.
(c,d) Interaction of two counter-propagating (c) or co-propagating (d) photons. \label{fig1}}
\end{center}
\end{figure}

%%%%%%%%%%%%%%%%%%%%%%%%%%%%%%%%%%%%%%%%%%%%%%%%%%%%

The basic physics is illustrated by considering 
a simple case [Fig.\ \ref{fig1}(b)], in which a single-photon wavepacket $\e$ propagates 
in an EIT medium [level scheme %shown 
in Fig.\ \ref{fig1}(a)] with a central control atom at $z=0$ prepared in a Rydberg state $|r^\prime\rangle$. Atoms in another Rydberg state  
$|r\rangle$, coupled by the EIT control laser  [Fig.\ \ref{fig1}(a)], experience a van der Waals potential $V(z) = C_6/z^6$ due to the interaction with the  
control atom, %at $z = 0$, prepared in a different Rydberg state $|r'\rangle$ 
which is 
decoupled from the 
applied %light 
fields. Alternatively, one could apply an electric field to induce dipole moments in states 
$|r\rangle$ and $|r'\rangle$, resulting in $V \propto 1/z^3$ \cite{saffman10}. 
%If $|r\rangle-|r\rangle$ interaction is negligible, the solution of this problem also applies to the propagation of a weak coherent pulse.

Sufficiently far away from $z = 0$, the incident photon propagates in a standard EIT medium featuring a 
two-photon-resonant control field with Rabi frequency $\Omega$. In the vicinity of the control atom, however, 
the state $|r\rangle$ is shifted so strongly out of resonance that the photon sees only a two-level ($|g\rangle$, $|e\rangle$) 
medium with transition linewidth $2 \gamma$.
The critical $z$, at which the interaction is equal to the EIT linewidth, separates these two regimes and corresponds to the Rydberg blockade
radius \cite{jaksch00,lukin01b}. 
%and , which In the present case, the blockade radius is set by the distance at which the interaction shift exceeds the EIT linewidth.
When the single-photon detuning $\Delta = 0$, the resonant blockade radius $z_\textrm{b}$ is thus defined by $V(z_\textrm{b}) = \Omega^2/\gamma$ ($\hbar = 1$), 
while for $\Delta \gg \gamma$, we define the off-resonant blockade radius $z_\textrm{B}$ via $V(z_\textrm{B}) = \Omega^2/\Delta$
(we assumed $\Delta/C_6  > 0$). 
Since the blockade region extends over $2 z_{\rm b(B)}$ % twice the blockade radius 
[Fig.\ \ref{fig1}(b)], the presence of the control atom
%induces an optical depth of 
locally creates an absorbing or refractive medium with an effective optical depth 
$d_{\rm b(B)} = 2 d z_{\rm b(B)}/L$, where $d$ is the resonant optical depth of the $|g\rangle-|e\rangle$ medium ($\Omega = 0$) of length $L$. 
%corresponding two-level medium ($\Omega=0$) with length $L$.
Interesting effects occur at large blockaded optical depths  $d_{\rm b(B)}$. 
In the resonant case, assuming $d_{\rm b} \gg 1$, the presence of the $|r'\rangle$
excitation causes complete scattering, i.e.\ loss, of the incoming photon.
Off resonance, for $d_\textrm{B} \gg 1$ and $d_\textrm{B} (\gamma/\Delta)^2 \ll 1$, %$L > 2 z_\textrm{b}$, 
the interaction with the $|r'\rangle$-atom 
imprints a phase $\sim d_\textrm{B} \gamma/\Delta$ on the probe photon and reduces its group delay by
$\sim d_\textrm{B} \gamma/\Omega^2$, as its group velocity is increased to the speed of light, $c$, within the blockade region. 

%In addition to developing the intuition for the multi-photon case, the above simple problem itself may be of practical use.
In the off-resonant case,  this simple system has direct practical applications.
First, by encoding a qubit in the ground and $|r'\rangle$ states of the central control atom, one 
can implement a phase gate between the probe photon and the atom.
 Second, the protocol of Ref.\ \cite{duan04c} allows to implement a phase gate between two
photons by successively sending them past the control atom that is appropriately prepared and manipulated 
between the passes. Selective manipulation of the control atom can be achieved particularly simply if it is of a different species or isotope. 
Third, a phase gate between two photons can also be achieved by storing one of them in
the $|r'\rangle$ state of the control atom and sending the other one through the medium. While storing
a single photon in a single atom is difficult, the same effect can be achieved by storing \cite{fleischhauer02,gorshkov07c}
the photon in a collective $|r'\rangle$ excitation, as we will discuss below. 

The results of this simple problem can be extended to the case of
%The intuition developed in this simple problem suggests the following results for 
multi-photon EIT propagation in 
Rydberg media. First, off-resonance, two counter-propagating photons [Fig.\ \ref{fig1}(c)]
can pick up a phase $\sim d_\textrm{B} \gamma/\Delta$, enabling the implementation
of a two-photon phase gate \cite{friedler05,shahmoon10}. %\cite{nielsen. 
%which has applications to efficient quantum computing and fast long-distance quantum  communication \cite{bouwmeester00,franson04}.
Second, a pulse of co-propagating photons [Fig.\ \ref{fig1}(d)] will evolve into a non-classical state corresponding to a train of 
single photons \cite{pohl10} and exhibiting correlations similar to those of hard-sphere particles with radius $z_{\rm b(B)}/2$.
These correlations arise from scattering of photon pairs within the blockade region. %slight cheating because they will also advance fast, but eventually they'll scatter unless the detuning is super high
Third, in the regime where $z_{\rm b}$ %$z_{\rm b(B)}$ 
is larger than the EIT-compressed pulse length, $\sigma$, both co- and
counter-propagating resonant setups might be usable as single-photon sources since all but one excitation will be extinguished. In the following, we present a detailed theoretical analysis of these phenomena.

\textit{Interaction of a photon with a stationary excitation.}---
We begin by detailing the solution of the %above 
problem of a single photon propagating in a medium where state $|r\rangle$ experiences a potential $V(z)$ [Fig.\ \ref{fig1}(b)].
Treating the medium in a one-dimensional continuum approximation, 
working in the dipole and rotating-wave approximations, and adiabatically eliminating the 
polarization on the $|g\rangle-|e\rangle$ transition, the slowly varying electric field amplitude $\e$ of the single-photon wavepacket and 
the polarization $S$ on the $|g\rangle-|r\rangle$ transition obey  \cite{fleischhauer02,gorshkov07c}
\begin{eqnarray}
&(\partial_t + c \partial_z) \e(z,t) = - \frac{g^2 n}{\Gamma} \e(z,t) - \frac{g \sqrt{n} \Omega}{\Gamma} S(z,t), \label{eq:e}\\
&\partial_t S(z,t) = - i U(z,t) - \frac{\Omega^2}{\Gamma} S(z,t) - \frac{g \sqrt{n} \Omega}{\Gamma} \e(z,t) \label{eq:s}.
\end{eqnarray}
Here $\Gamma = \gamma - i \Delta$, $U(z,t) = V(z) S(z,t)$, $g$ is the atom-field coupling constant, and $n$ is the 
atomic density. We have neglected the depletion of state $|g\rangle$ and the finite lifetime of the Rydberg state $|r\rangle$, which 
is typically much longer than the propagation times considered here \cite{saffman10}.
Assuming that all atoms are in state $|g\rangle$ before the arrival of the photon, Eqs.\ (\ref{eq:e},\ref{eq:s}) can be solved
to give
\begin{eqnarray}
\e\left(\tfrac{L}{2},t\right) \!=\! \int_{-\infty}^{\infty} d \omega e^{- i \omega \left(t - \frac{L}{c}\right)+ i \tfrac{k}{2}
\int_{-\frac{L}{2}}^\frac{L}{2} d z \chi(z,\omega)} \tilde \e\left(-\tfrac{L}{2},\omega\right), \label{eq:ft}
\end{eqnarray}
where 
\begin{equation}
k \chi(z,\omega) = \frac{1}{L} \frac{d \gamma [\omega - V(z)]}{\Omega^2 - (\Delta + i \gamma) [\omega - V(z)]} \label{eq:fdef}
\end{equation}
and $\tilde \e\left(-L/2,\omega\right)$ is the Fourier transform of the wavepacket incident at $z = -L/2$.

 %and $V$ of the same sign. 
For narrowband pulses, we expand $\chi$ in $\omega$ and, assuming %and carry out the integrations in Eq.\ (\ref{eq:ft}). For 
$\Delta \gg \gamma$ and $L \gg 2 z_\textrm{B}$, reduce Eq.\ (\ref{eq:ft}) to
%Off resonance ($\Delta \gg \gamma$), 
%this gives
\begin{equation}
\e\left(L/2,t\right) \approx \e\left(-L/2,t-L^{\prime}/v_{\rm g}\right)e^{i\varphi-\eta},
\end{equation}
where $v_{\rm g}\approx c\Omega^2/(g^2n) = 2 \Omega^2 L/(d \gamma)$ is the EIT group veclocity. In order to avoid the Raman resonance at $V + \Omega^2/\Delta = 0$, we assumed $\Delta/C_6 > 0$. %reduced group velocity inside the EIT medium.
Since the photon travels at $c$ within the blockade region, the group delay comes from  %the interactions result in a modified group delay, described by 
a reduced medium length $L^{\prime}=L-\tfrac{7}{9}\pi z_{\rm B}\approx L-2z_{\rm B}$. Additionally, the emergence of a two-level medium within $|z|<z_{\rm B}$ 
gives an intensity attenuation of $e^{-2\eta}$ with $2\eta=\tfrac{5\pi}{18}d_{\rm B}(\gamma/\Delta)^2\approx d_{\rm B}(\gamma/\Delta)^2$ and a picked-up phase of $\varphi=-\tfrac{\pi}{6}d_{\rm B}(\gamma/\Delta)\approx -\tfrac{1}{2}d_{\rm B}(\gamma/\Delta)$.
Thus, with $d_\textrm{B} \gg 1$ and a properly chosen $\Delta \gg \gamma$, one can get a considerable phase and/or change in group delay without significant absorption. 
For the same derivation on resonance ($\Delta = 0$), the main effect is an intensity attenuation of  
%$\exp\left[-\tfrac{1}{12} \Gamma\left(\frac{1}{12}\right) \Gamma\left(\frac{11}{12}\right) d_\textrm{b}\right]$ 
$\approx \exp(-d_\textrm{b})$, as expected for a two-level medium of length $2z_{\rm b}$.

It is straightforward to extend our analysis to a delocalized $|r'\rangle$ excitation, i.e.\ a spin wave, that is spread over many atoms. Far off resonance, the effect of the control atom is independent of its position, such that a single control atom and a  corresponding spin wave affect the incident photon identically.  On resonance, with $d_\textrm{b} \gg 1$, the $|r'\rangle$ spin wave causes complete scattering of the incoming photon. At the same time, after tracing out the scattered photon, which carries information about the location of the scattering, the spin wave itself disentangles into a classical mixture of pieces of length $\sim z_\textrm{b}$.

\textit{Interaction of propagating photons.}---We now consider the problem of propagating photons interacting with each other.
Regarding $\e$ and $S$ in Eqs.\ (\ref{eq:e},\ref{eq:s}) as operators with same-time commutation relations
 $[\e(z),\e^\dagger(z')] = [S(z),S^\dagger(z')] = \delta(z-z')$  \cite{gorshkov07c} %neglecting Langevin noise, 
 and taking $U(z) = \int d z' V(z-z') S^\dagger(z') S(z') S(z)$, Eqs.\ (\ref{eq:e},\ref{eq:s}) become Heisenberg operator equations \cite{langevin}
% \footnote{Langevin noise does affect our calculations.} 
 for the case of photons co-propagating in a
Rydberg EIT medium [Fig.\ \ref{fig1}(d)]. Alternatively, for the case of two counter-propagating photons [Fig.\ \ref{fig1}(c)], 
we define operators $\e_{1(2)}$ and $S_{1(2)}$ for the right- (left-)moving photon. For $S_1$, the interaction is 
$U(z) = \int d z' V(z-z') S_2^\dagger(z') S_2(z') S_1(z)$, and vice versa for $S_2$.

Since the physics of two counter-propagating photons is similar to the spin-wave problem above, %photon interaction with a spin wave, 
we begin our analysis with this case [Fig.\ \ref{fig1}(c)]. 
Letting $|\psi(t)\rangle$ be the two-excitation
wavefunction \cite{hafezi09b}, we define
$ee(z_1,z_2,t) = \langle 0| \e_1(z_1) \e_2(z_2) |\psi(t)\rangle$, $es(z_1,z_2,t) = \langle 0| \e_1(z_1) S_2(z_2) |\psi(t)\rangle$,
 $se(z_1,z_2,t) = \langle 0| S_1(z_1) \e_2(z_2) |\psi(t)\rangle$, 
 and $ss(z_1,z_2,t) = \langle 0| S_1 (z_1) S_2 (z_2) |\psi(t)\rangle$.
 %and $ss(z_1,z_2,t) \!=\! \langle 0| S_1\!(z_1) S_2\!(z_2) |\psi(t)\rangle$.
Eqs.\ (\ref{eq:e},\ref{eq:s}) then yield a system of equations for these four variables.
Defining $es_\pm = (es \pm se)/2$, one finds that $es_-$ is small and does not significantly affect the dynamics.
Dropping $es_-$, defining center-of-mass and relative coordinates $R = (z_1 + z_2)/2$ and $r = z_1-z_2$, and taking a
Fourier transform in time, one obtains $c \partial_r \mathbf{v} = \mathbf{M}(r,\omega) \mathbf{v}$, where $\mathbf{v} = \{ee(R,r,\omega), es_+(R,r,\omega)\}$ and
\begin{eqnarray}
\mathbf{M}(r,\omega) = \left[
\begin{array}{cc}
i \frac{w}{2} - \frac{g^2 n}{\Gamma}  & - \frac{g \sqrt{n} \Omega}{\Gamma} \\
-\frac{g \sqrt{n} \Omega}{\Gamma} & i \omega- \frac{\Omega^2}{\Gamma} +
 \frac{  i g^2 n [\omega - V(r)]}{2 \Omega^2 + i V(r) \Gamma - i \omega \Gamma}
\end{array}
\right].
\end{eqnarray}
$R$ enters only through boundary conditions and is, thus, not important in the present case. For narrowband pulses, we can expand $\mathbf{M}(r,\omega) \approx \mathbf{M}_0(r) + \omega \mathbf{M}_1(r)$, with
\begin{subequations}\label{eq:M}
\begin{eqnarray}
\label{eq:M0}
\mathbf{M}_0 &=& - \frac{1}{\Gamma} \left[
\begin{array}{cc}
 g^2 n  &  g \sqrt{n} \Omega \\
g \sqrt{n} \Omega & \Omega^2 +g^2 n \mathcal{V}
\end{array}
\right]\;,\\
\label{eq:M1}
\mathbf{M}_1 &=&  i \left[
\begin{array}{cc}
 \tfrac{1}{2}  &  0 \\
0 & 1-2g^2n\tfrac{\Omega^2\mathcal{V}^2}{\Gamma^2V^2}
\end{array}
\right]\;. 
\end{eqnarray}
\end{subequations}
Here we defined the effective potential $\mathcal{V}=\Gamma V/(\Gamma V-i2\Omega^2)$. Outside (inside) the blockade region, $\mathcal{V} \approx i \Gamma V/(2 \Omega^2)$ ($\mathcal{V} \approx 1$).
 %$\propto(z_{\rm b(B)}/r)^6$ 
% while $\mathcal{V}(r \rightarrow 0) = 1$. %within the blockade radius 
%saturates to $1$ as $r\rightarrow0$.
%Outside the blockade radius ($V(r) \approx 0$), 
For $|r| \gg z_\textrm{b(B)}$, the two photons propagate as dark-state polaritons \cite{fleischhauer02}, i.e.\ we have $es_+/ee = - g \sqrt{n}/\Omega$, which is an eigenstate of $\mathbf{M}_0$ with eigenvalue $0$.
Since $g \sqrt{n} \gg \Omega$, the group velocity can be read out from the last entry of $\mathbf{M}_1$,
which %, as expected, 
gives twice the EIT group velocity $v_g$ %= c \Omega^2/(g^2 n)$ 
since the two 
polaritons propagate towards each other. Within the blockade radius, where $\mathcal{V}\approx1$ and $\mathcal{V}/V\approx0$, the polariton solution ceases to be an eigenstate of $\mathbf{M}_0$, and Eq.\ (\ref{eq:M1}) predicts a speed up to $\sim c$. Since the time %$t_{\rm b(B)}\sim$ 
$\sim z_{\rm b(B)}/c$ it takes to cross the blockade region is much less than the inverse width of the EIT window, the dynamics is highly non-adiabatic such that the main result of the interactions is a picked-up factor of $\exp\left[-\int dr g^2 n \mathcal{V}(r)/(c \Gamma)\right] = \exp(i\varphi-\eta)$. This is a generalization of the result of Refs.\ \cite{friedler05,shahmoon10} beyond the perturbative regime.

 %$\exp\left[-\int dr \tfrac{g^2 n}{c\Gamma} \mathcal{V}(r)\right]$.

On resonance, %this gives an intensity attenuation by 
$2 \eta \approx d_\textrm{b}$. %$\approx \exp(-d_\textrm{b})$.
%$\exp[ - \pi 3^{-1} 2^{-2/3} (\sqrt{3}-1)^{-1} d_\textrm{b}] \approx \exp(-d_\textrm{b})$. 
%$\exp(-\tfrac{\pi}{2^{1/6}3}d_\textrm{b})\approx\exp(-d_\textrm{b})$ . 
Thus, analogously to the spin-wave problem above, the entire EIT-compressed two-particle wavefunction decays provided %if the EIT-compressed two-particle wavefunction 
it fits inside the medium and %provided 
$d_\textrm{b} \gg 1$. 
The resulting state is a statistical mixture of right- and left-moving excitations.

Off resonance, %integration shows that after passing the blockade region, 
$es_+$ %has 
picks up %ed up a phase factor $e^{i\varphi-\eta}$, where  
 $\varphi \approx -\tfrac{\pi}{2^{1/6}6}\tfrac{\gamma}{\Delta}d_{\rm B}\approx-\tfrac{\gamma}{2\Delta}d_{\rm B}$ and $\eta \approx \tfrac{5\pi}{2^{1/6}36}\tfrac{\gamma^2}{\Delta^2}d_{\rm B}\approx\tfrac{\gamma^2}{2\Delta^2}d_{\rm B}$. Additionally, the off-diagonal terms in $\mathbf{M}_0$ result in a small admixture of the bright-state polariton \cite{fleischhauer02}, which decays after the wavefunction exits the blockade region. %and produces some additional loss. 
 
 To verify these conclusions, we show in Fig.\ \ref{fig:counter} and in the supplementary movie \cite{movies} the results of numerical solutions of the full equations for $ee$, $es$, $se$, and $ee$ in the off-resonant case. 
%The damped oscillations following the interaction region indicate the admixture and decay of the bright-state-polariton. 
Despite the bright-polariton-induced %strong 
oscillations of $ee$ inside and near the blockade region \cite{movies}, the final phase of the outgoing two-photon pulse perfectly 
agrees with our analytical prediction [Fig.\ \ref{fig:counter}(e)]. 
While also showing good agreement with the analytical result, the obtained loss 
%The numerically obtained loss 
is slightly larger due to the bright-state polariton admixture, which was neglected within the above approximate treatment. 
%For the parameters of interest ($d_{\rm B}\sim1$), the analytical formula, however, yields a good estimate of the actual photon loss.

Provided the EIT-compressed two-particle wavefunction fits inside the medium, this process, thus, allows for the implementation of a nearly lossless phase gate between two photons. Taking a specific example of cold Rb atoms with $|e\rangle = 5{}^2P_{1/2}$ and $|r\rangle = 70{}^2S_{1/2}$ and using $\Omega/2\pi=2$MHz and $\Delta=20\gamma$, we find $z_{\rm B} = 15 \mu$m, which, for a dense cloud with  $n = 10^{12} \textrm{ cm}^{-3}$,  %and $\lambda = 795$ nm, 
gives $d_\textrm{B} = \frac{3}{2 \pi} \lambda^2 (2 z_\textrm{B}) n \approx 9$. This yields a significant phase of $\varphi\approx-0.2$ and a very small attenuation $2 \eta \approx 0.02$. %$e^{-0.02}$. 
One can  increase $d_\textrm{B}$ further by using 
photonic waveguides \cite{christensen08,bajcsy09,vetsch10,kohnen11} and working with a BEC \cite{kohnen11}.

\begin{figure}[t!]
\begin{center}
\includegraphics[width = 0.99 \columnwidth]{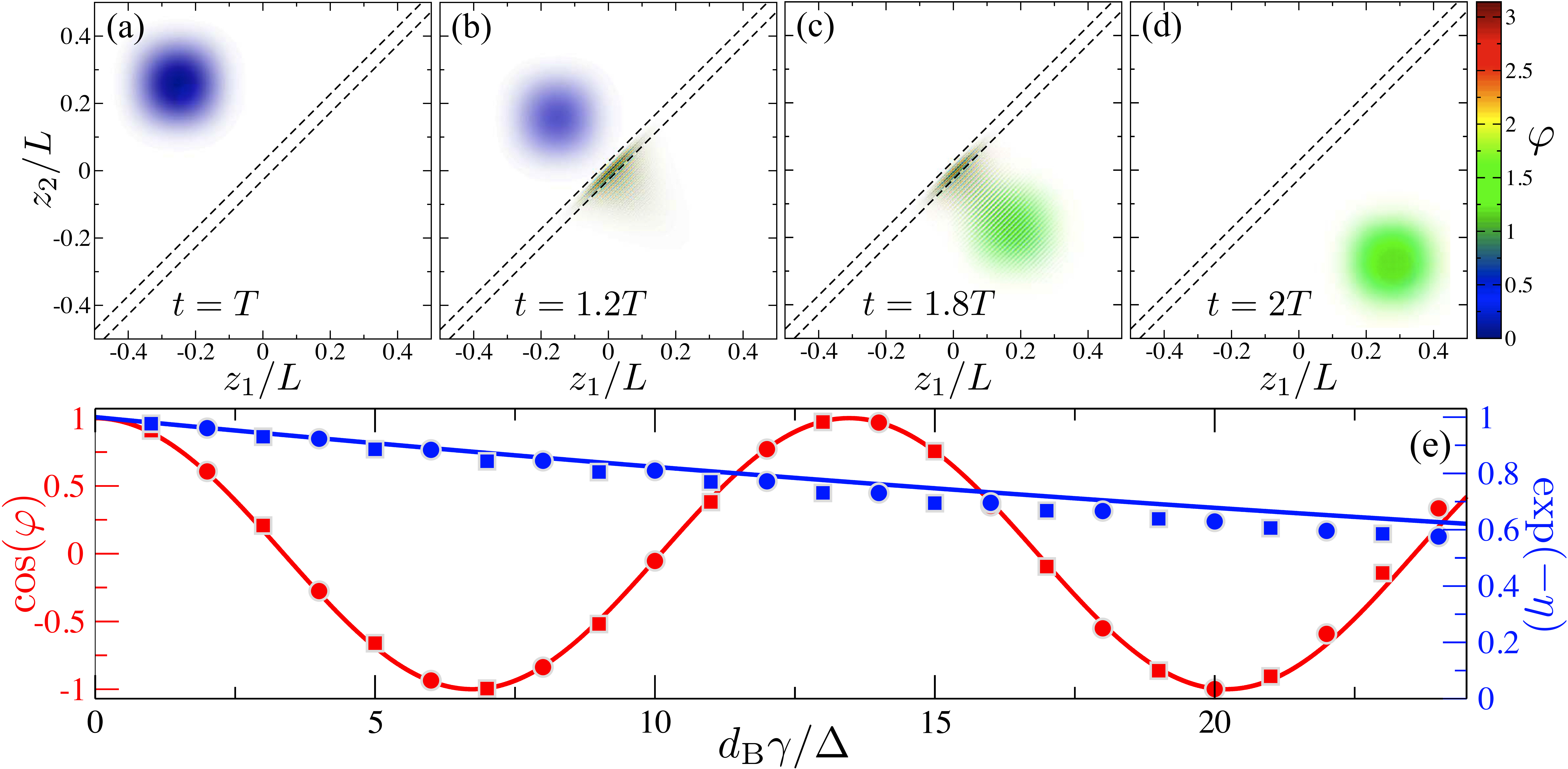}
\caption{(a)-(d) Two-photon counter-propagation for $\Delta=20\gamma$, $\Omega=2\Delta$, $g \sqrt{n}=20\Delta$ and $z_{\rm B}=0.055\sigma$, where $\sigma$ is the compressed pulse length inside the medium. The color coding shows the local phase of $ee$, while the opacity reflects the two-photon density $|ee|^2$. The dashed lines are $|z_1 - z_2| = z_\textrm{B}$. The full movie is provided in the supplementary material \cite{movies}. (e) Numerically obtained  phase shift $\varphi$ [we plot $\cos \varphi$] and attenuation $e^{-\eta}$ as a function of $d_{\rm B}$ compared to the analytical predictions (lines). The numerical data corresponds to two different parameter scans $g\sqrt{n}=400\Delta$, $z_{\rm B}=0.0025\sigma,...,0.03\sigma$ (dots) and $z_{\rm B}=0.03\sigma$, $g \sqrt{n}=80,...,390$ (squares).  \label{fig:counter}}
\end{center}
\end{figure}

%\begin{video}
%\includegraphics[width = 0.99 \columnwidth]{fig2.pdf}
%\setfloatlink{http://some.video.com/fun.mov}
%\caption{\label{vid:interest}This is a video
%of something fun. This is a video
%of something fun.}
%\end{video}

In the co-propagating case, we define $ee(z_1,z_2,t)  = \langle 0| \e(z_1) \e(z_2) |\psi(t)\rangle$, $es(z_1,z_2,t) = \langle 0| \e(z_1) S(z_2) |\psi(t)\rangle$, and $ss(z_1,z_2,t) = \langle 0| S(z_1) S(z_2) |\psi(t)\rangle$ [Fig.\ \ref{fig1}(d)]. Defining $es_\pm(z_1,z_2) = [es(z_1,z_2) \pm es(z_2,z_1)]/2$, dropping $es_-$, and taking the Fourier transform in time, we obtain $c \partial_R \mathbf{v} = 2 \mathbf{M}(r,\omega) \mathbf{v}$. That is, the only difference from the counter-propagating case is the replacement of $\partial_r$ with $(1/2) \partial_R$. The resulting equations can be solved separately at each $r$. 
As before, outside the blockade radius, the two-photon dark-state polariton propagates with group velocity $v_g$. 
Inside the blockade radius, $\mathbf{M}_0$ results in fast attenuation on a lengthscale $ \sim \tfrac{L}{d}(\gamma^2+\Delta^2)/\gamma^2$.   
This is confirmed by our numerical calculations, shown in Fig.\ \ref{fig:co} and in the supplementary movie \cite{movies}. 
Therefore, the two-excitation wavefunction evolves into a statistical mixture of a single excitation and a correlated train of two photons separated by $z_\textrm{b(B)}$. On resonance, if the photon is scattered when the EIT-compressed pulse of length less than $z_\textrm{b}$ is fully inside the medium, we expect the remaining excitation to propagate in its original spatiotemporal mode.  
For a coherent  input pulse, one similarly expects the wavepacket to evolve with some probability into a correlated train of blockade-radius-separated 
 photons. Furthermore, if the blockade radius is larger than the EIT-compressed pulse length, there may be a regime in which such a system could function as a deterministic single-photon source.
\begin{figure}[t!]
\begin{center}
\includegraphics[width = 0.99 \columnwidth]{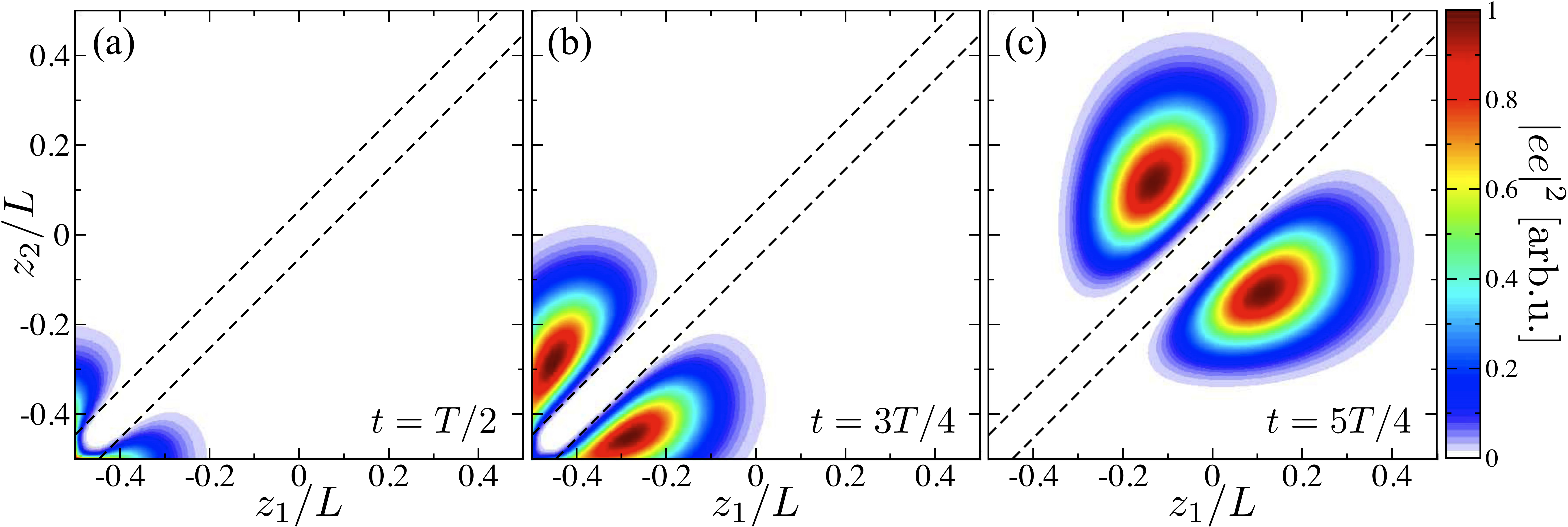}
\caption{Time evolution of $|ee|^2$ for two co-propagating photons for $\Omega=\gamma$, $g \sqrt{n}=100\gamma$, and $z_{\rm b}=0.08\sigma$. The dashed lines are $|z_1 - z_2| = z_\textrm{b}$, in agreement with the numerical results, which show the decay of $ee$ within the dashed lines. The full movie is provided in the supplementary material \cite{movies}. \label{fig:co}}
\end{center}
\end{figure}

%\textit{Conclusion.}---
In summary, we have shown that Rydberg blockade in EIT media can be harnessed for inducing strong photon-photon interactions, with applications to generating nonclassical states of light, implementing nonlinear photonic gates, and studying many-body phenomena with strongly correlated light. This work opens several promising avenues of research. With an eye towards single-photon generation, one can extend the presented wavefunction treatment to a density matrix approach and explicitly analyze the propagation of the remaining excitation after the interaction-induced decay of multi-photon states. In addition, a gas of bosons (Rydberg polaritons) with a hard-sphere core (of radius $z_\textrm{b(B)}/2$) 
can be investigated both theoretically and experimentally
in the co-propagating case. %after those photons that are within $z_\textrm{b(B)}$ of each other have decayed or advanced. 
In particular, the previously neglected effects of $es_-$ %that we have neglected 
 endow these bosons with an effective mass $\propto -i d \gamma/(L v_g \Gamma)$, 
%(a term $\propto \partial_r^2$), 
which %may 
plays a significant role for propagation distances %much 
larger than those considered in the present Letter. By including the effects of the coordinates transverse to the propagation axis, one can extend this problem to higher dimensions. Furthermore, for $\Delta/C_6 < 0$, the effective potential shows a resonant feature, which can give rise to 
two-polariton bound states. 

We thank S.\ Hofferberth, M.\ Bajcsy, T.\ Peyronel, M.\ Hafezi, and N.\ Yao for discussions. This work was supported by NSF, %(Grants No. PHY-0803371, %IQI
%PHY05-51164), %KITP
 the Lee A. DuBridge Fellowship, DFG through the GRK 792, CUA, DARPA, and the Packard Foundation.


\begin{thebibliography}{38}
\bibitem{langevin} Additional Langevin noise does not affect our calculations.
\end{thebibliography}
\end{document}